\documentclass[manuscript,screen]{acmart}
\AtBeginDocument{%
  \providecommand\BibTeX{{%
    \normalfont B\kern-0.5em{\scshape i\kern-0.25em b}\kern-0.8em\TeX}}}

\setcopyright{acmcopyright}
\copyrightyear{2022}
\acmYear{2022}
\acmDOI{XXXXXXX.XXXXXXX}

\acmConference[CHI '22 Workshop Paper]{CHI '22: The 2022 ACM CHI Conference on Human Factors in Computing Systems}{April 30 -- May 6, 2022}{New Orleans, LA, USA}
\acmBooktitle{CHI '22: The 2022 ACM CHI Conference on Human Factors in Computing Systems}
\acmPrice{15.00}
\acmISBN{978-1-4503-XXXX-X/18/06}

\usepackage{soul}
\usepackage{makecell}
\usepackage{subfigure}

\begin{document}

\title[Making Hidden Bias Visible]{Making Hidden Bias Visible: Designing a Feedback Ecosystem for Primary Care Providers}

\author{Naba Rizvi}
\affiliation{%
  \institution{University of California, San Diego}
  \city{La Jolla, CA}
  \country{United States}
}

\author{Harshini Ramaswamy}
\affiliation{%
  \institution{University of California, San Diego}
  \city{La Jolla, CA}
  \country{United States}
}

\author{Reggie Casanova-Perez}
\affiliation{%
  \institution{University of Washington}
  \city{Seattle, WA}
  \country{United States}
}

\author{Andrea Hartzler}
\affiliation{%
  \institution{University of Washington}
  \city{Seattle, WA}
  \country{United States}
}

\author{Nadir Weibel}
\affiliation{%
  \institution{University of California, San Diego}
  \city{La Jolla, CA}
  \country{United States}
}

\renewcommand{\shortauthors}{Rizvi et al.}

\begin{abstract}
Implicit bias may perpetuate healthcare disparities for marginalized patient populations. Such bias is expressed in communication between patients and their providers. We design an ecosystem with guidance from providers to make this bias explicit in patient-provider communication. Our end users are providers seeking to improve their quality of care for patients who are Black, Indigenous, People of Color~(BIPOC) and/or Lesbian, Gay, Bisexual, Transgender, and Queer (LGBTQ). We present wireframes displaying communication metrics that negatively impact patient-centered care divided into the following categories: digital nudge, dashboard, and guided reflection. Our wireframes provide quantitative, real-time, and conversational feedback promoting provider reflection on their interactions with patients. 
Through a design critique, we found primary care providers prefer technologies that are efficient, context-aware, private, and address barriers. This is the first design iteration toward the development of a tool to raise providers’ awareness of their own implicit biases.
\end{abstract}



%\keywords{datasets, neural networks, gaze detection, text tagging}


\maketitle

\vspace{-0.1cm}
\section{Introduction}
Biases in healthcare professionals are of great concern because they can lead to disparities in quality of care and poorer patient health outcomes \cite{cooper2012associations}. Unlike explicit bias, implicit bias is unintentional or unconscious, causing challenges in identifying and mitigating it~\cite{blair2011unconscious}.
%
%which can make identifying and mitigating it quite difficult \cite{blair2011unconscious}.
%
Implicit biases contribute to racial disparities in healthcare by influencing patient-provider interactions, treatment decisions, and patient health outcomes~\cite{hagiwara2019detecting, fitzgerald2017implicit}. In a systematic review of implicit bias and disparities in healthcare outcomes~\cite{maina2018decade}, six studies showed that providers with high implicit bias measured using the Implicit Association Test (IAT) exhibited poorer communication with patients.  
 
%Our study
We build upon prior work that has proposed frameworks and highlighted opportunities in designing provider-facing systems to mitigate implicit bias in healthcare \cite{sukhera2020implicit, dirks2022, loomis2021human}. 
In recent years, there has been a growing interest in health equity dashboards to improve patient health outcomes \cite{blagev2021journey, cookson2018health, tsuchida2021developing, holeman2020human}. A review of clinical dashboards highlighted that while the implementation of dashboards that provide quick access to information to providers may help improve patient health outcomes, more research needs to be done to establish guidelines for the designs of such technology \cite{dowding2015dashboards}. 
One study found a significant association between difficult or poor user interface design of similar technologies and frustration levels which may lead to burnout among providers \cite{khairat2019mixed}. 
To address these gaps in usability research of provider-facing technologies, we present preliminary results of a design critique highlighting design considerations for feedback technologies that may mitigate implicit bias. We are employing a user-centered design approach \cite{norman1986user} in our study to incorporate feedback from providers in our work through iterative critiques with primary care providers as our target users. The main research question guiding our study is:  What are the design considerations for a provider-facing ecosystem addressing implicit bias in patient-provider communication?

\vspace{-0.1cm}
\section{Related Work}
\textbf{Social Signal Processing.} Collecting nonverbal communication cues can help improve the quality of feedback that providers receive on their interpersonal communication skills, which may ultimately improve health outcomes for patients~\cite{ha2010doctor}. Specific guidelines for delivering feedback to providers recommend including data in timely customizable reports that highlight patterns and include glanceable summaries~\cite{mcnamara2016confidential}. 
Our work is novel in applying Social Signal Processing (SSP) to design a feedback ecosystem for patient-provider interactions. SSP is a computational approach for extracting social signals such as turn-taking and eye contact in human interactions~\cite{vinciarelli2009social}, and we use it to obtain feedback on communication patterns.
While prior research has demonstrated clinician acceptability of SSP feedback on patient-provider communication~\cite{patel2013visual}, delivered such feedback in online interactions \cite{faucett2017should}, and to medical students \cite{liu2016eqclinic}, providers' perspectives on a multimodal ecosystem highlighting their implicit bias in patient-provider communication has not yet been investigated. 
%
%Prior work has demonstrated clinician acceptability of SSP feedback on patient-provider communication~\cite{patel2013visual}. However, prior research has not investigated clinicians' perspectives on the design of SSP feedback on implicit bias in patient-provider communication~\cite{hartzler2014real}. 
%
In this work, we present designs of SSP-powered feedback systems to make implicit bias explicit to providers. The metrics visualized in our wireframes were selected based on prior work on ambient feedback from SSP models to facilitate patient-centered communication~\cite{hartzler2014real} and other works on behaviors affiliated with implicit bias~\cite{fitzgerald2017implicit,patel2013visual,jaramillo2021guidance,kanter2020addressing}. 

\noindent \textbf{User-Centered Design. }Previous work has shown the significance of incorporating feedback into the design process to assess whether the design is becoming more effective for end users~\cite{easterday2014computer}. We are engaging primary care providers in iterative semi-structured design critiques so our ecosystem can better serve their needs. Our work builds upon an existing %previous
needs assessment of providers that contributed three design recommendations to identify and mitigate implicit bias during provider-patient interactions \cite{dirks2022}. The first recommendation is a digital ``nudge'', such as an alert on a digital device notifying the provider of changes in their communication behaviors with patients during the appointment. The second recommendation is a reflective tool to help clinicians process feedback on implicit bias with another person or group. The third recommendation is a data-driven feedback system, such as a dashboard that can visualize quantitative data about communication patterns across patients. We designed wireframes of an ecosystem following these recommendations and obtained feedback from end users. 

\section{Methods}
Our study was divided into three parts: 1) designing provider-facing feedback systems, 2) interviewing providers to obtain feedback and refine our designs, and 3) creating an affinity map summarizing their design critique. 

\subsection{Design}
\begin{figure}[h!]
    \vspace{-.75em}
    \centering
    \subfigure{\includegraphics[width=0.3\textwidth]{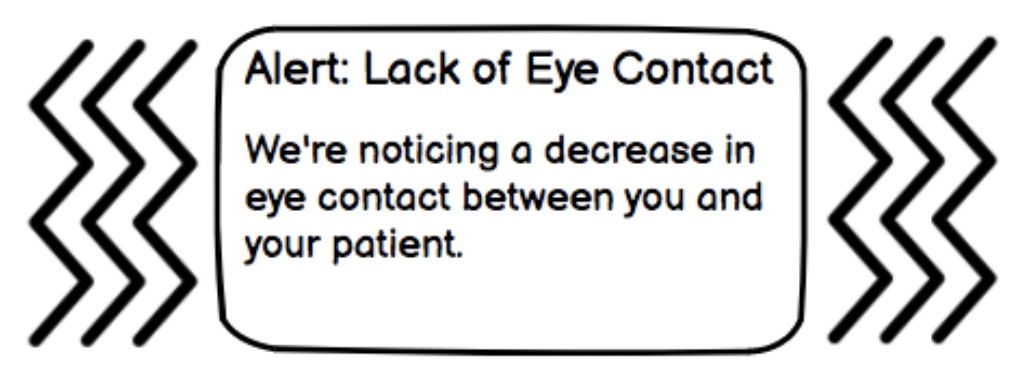}} 
    \subfigure{\includegraphics[width=0.3\textwidth]{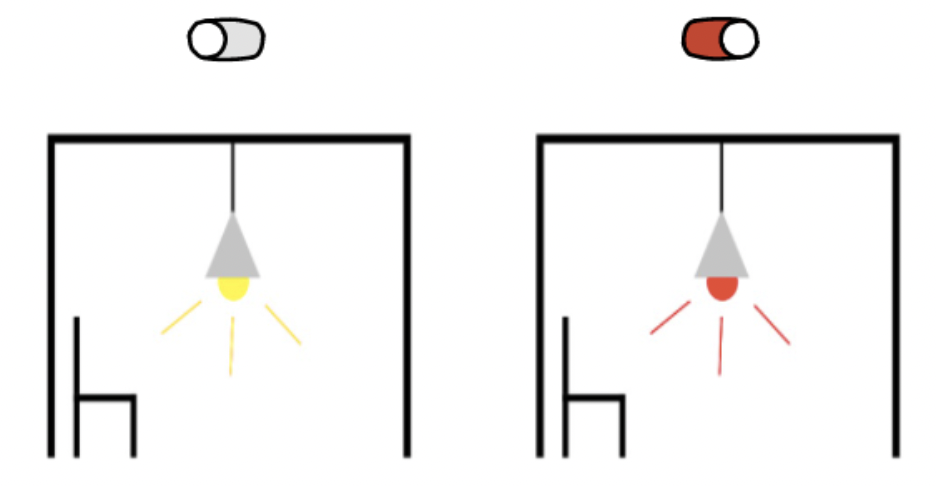}}
    \vspace{-.5em}
    \caption{Our digital nudge wireframes provide real-time feedback to providers using smartwatch notifcations (left) and changes in ambient lighting in the room (right).}
    \label{fig:digitalnudge}
    \vspace{-.75em}
\end{figure}

\begin{figure}[t!]
    \centering
    \fbox{
    \subfigure{\includegraphics[width=0.28\textwidth]{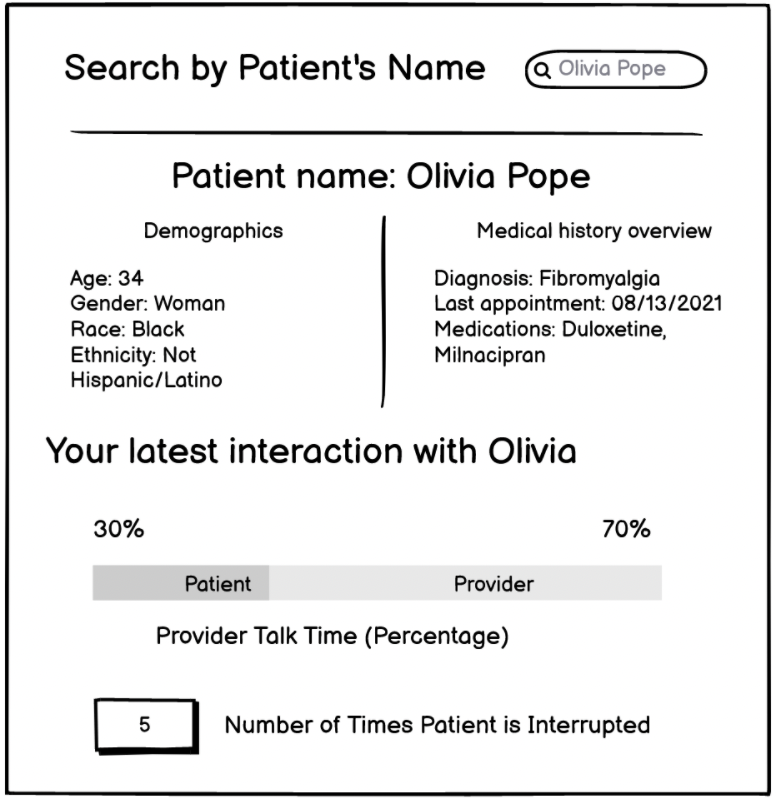}} 
    \subfigure{\includegraphics[width=0.345\textwidth]{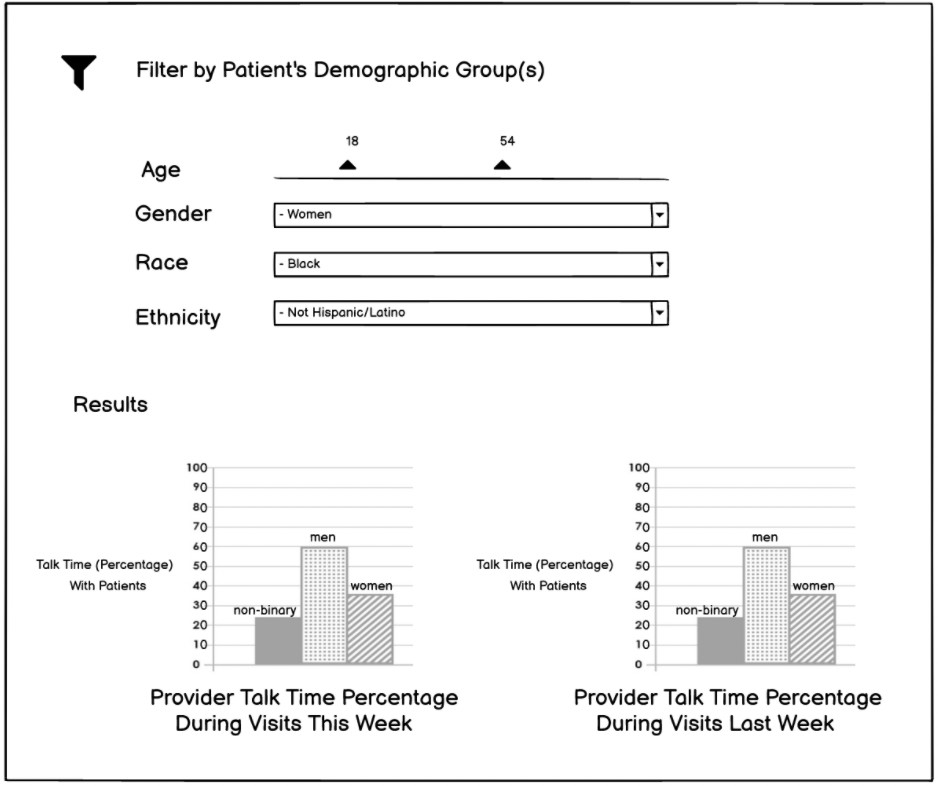}}
    }
    \subfigure{\includegraphics[width=0.3\textwidth]{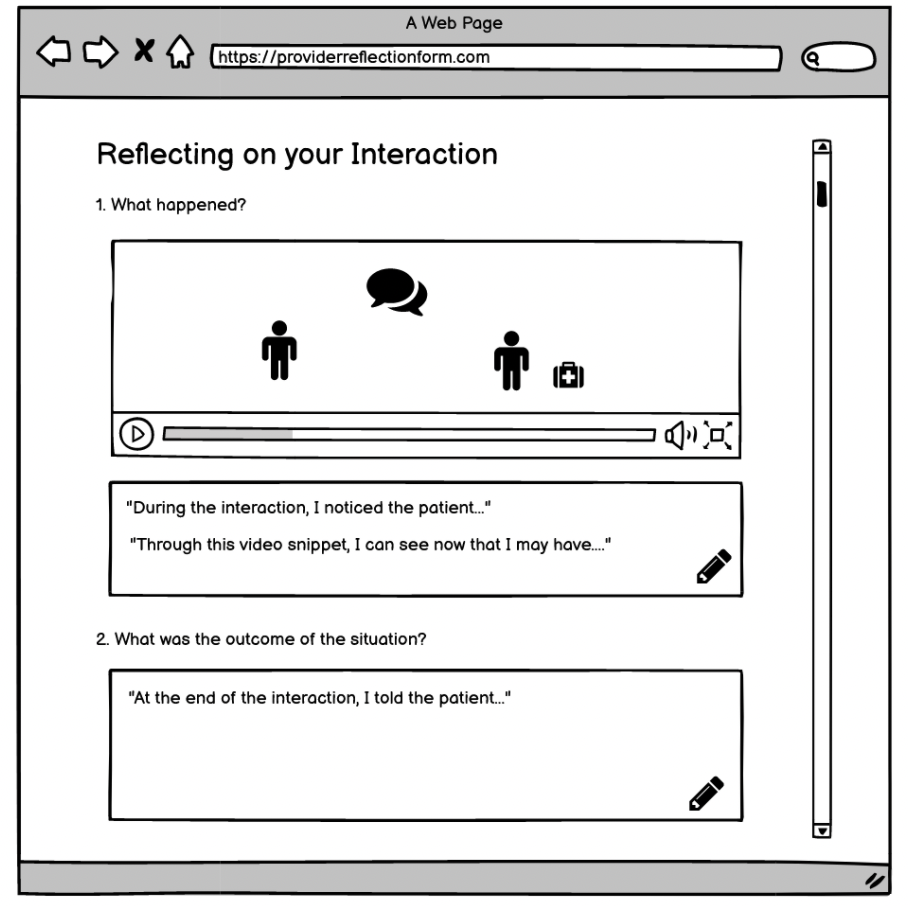}}
    \vspace{-.5em}
    \caption{\textbf{Left:} Dashboard wireframes visualizing communication metrics that impact patient-centered communication. The first version displays metrics from a single clinical interaction. The second version displays comparative metrics from weekly interactions based on patient demographics. \textbf{Right:} Guided reflection wireframe facilitating introspection and discussion among providers on their clinical interactions with patients. This can be in-person with a trusted third party or self-guided (pictured) through a digital form.}
    \label{fig:dashboard+reflection}
\end{figure}

We designed six multi-modal low-fidelity wireframes using Balsamiq software for rapid wireframing~\cite{balsamiq2015rapid}. A brief explanation of the metrics and design are explained in Table \ref{tab:Wireframes} and examples of the wireframes are shown in Figures \ref{fig:digitalnudge} and \ref{fig:dashboard+reflection}. We selected communication metrics based on studies discussing behaviors affiliated with implicit bias~\cite{patel2013visual,fitzgerald2017implicit, jaramillo2021guidance, kanter2020addressing, hartzler2014real} and explored ways of visualizing these metrics following the recommendations presented in a previous study \cite{dirks2022}. The metrics used in our study include: provider verbal dominance (talk time), interruptions, and proportion of eye contact between patients and providers during a visit. We conducted three pilot sessions eliciting design feedback from team members specializing in biomedical informatics and human-centered design in preparation for our design critique with providers.

\begin{table}
\centering
\begin{tabular}{|l|l|l|} 
\hline
\textbf{Wireframe} & \textbf{Metrics and Frameworks Used}                                                                                                                                               & \textbf{Design Rationale}                                                                                \\ 
\hline
Dashboard          & Verbal dominance and interruptions \text{\cite{patel2013visual, hartzler2014real}}                                         & Data-driven feedback \text{\cite{dirks2022}}                                                   \\ 
\hline
Digital Nudge      & Eye Contact and tone of voice \text{\cite{jaramillo2021guidance, fitzgerald2017implicit,hartzler2014real}} & \makecell[cl]{Real-time feedback on patient-centered \\ communication \text{\cite{dirks2022}}}  \\ 
\hline
Guided Reflection  & \makecell[cl]{A reflection on providers' communication\\ behaviors that may indicate implicit bias \\using the Gibbs Reflective Cycle. \text{\cite{gibbs1988learning}}  }                                                                                        & \makecell[cl]{Discussing and processing implicit bias\\ with a human \text{\cite{dirks2022}}}             \\
\hline
\end{tabular}
\caption{A brief overview of our wireframes and their design rationale.}
\label{tab:Wireframes}
\end{table}

\subsection{Interviews and Feedback}

\textbf{Participants and Recruitment.}
We conducted 1-hour semi-structured interviews on Zoom with six primary care providers from two large academic healthcare systems on the West Coast of the United States. The majority of the participants were white (66.7\%) and men (66.7\%), while 33.3\% of our participants were women. Comparatively, 72.5\% of primary care providers in the United States are white and 46.9\% are women~\cite{xierali2018racial}. 50\% of our participants had under 10 years of experience while the remainder had over 30 years of experience.
%\begin{table}[h!]
%\vspace{-.1cm}
%\begin{small}
%\begin{tabular}{|l|l|l|l|l|}
% \hline
% \textbf{Provider ID} & \textbf{Age} & \textbf{Race}         & \textbf{Gender}         & \textbf{Years in Practice} \\ \hline
% P1          & 40s & Latinx       & Man            & 9                \\ \hline
% P2          & 30s & Asian Indian & Man, Cisgender & 8                \\ \hline
% P3          & 30s & White        & Woman          & 8                \\ \hline
% P4          & 60s & White        & Man            & 41               \\ \hline
% P5          & 60s & White        & Woman          & 42               \\ \hline
% P6          & 50s & White        & Man            & 31               \\ \hline
% \end{tabular}
% \end{small}
% \caption{Summary of characteristics among healthcare providers who participated in the study (n=6).}
% \label{tab:demographics}
% \vspace{-.6cm}
% \end{table}

\noindent\textbf{Semi-Structured Interviews, and Wireframes Feedback.}
Our interviews commenced with a five minute introduction, followed by a semi-structured list of questions on the following topics: features, general use, timing of interaction with our ecosystem (e.g. before, after, or during patient visits), and confidentiality of the metrics shared. The wireframes were displayed in a slideshow over Zoom, and feedback was obtained for all prototypes. The sessions were recorded and the interviewers filled out a template containing field notes on the participant’s responses about the wireframes shown, new ideas generated by them, implications of using the feedback systems in their workflow, current methods of assessing communication with patients, and institutional supplementary materials that could help make the feedback ecosystem more feasible. 

\subsection{Analysis} 
We conducted affinity mapping to characterize design considerations based on participants’ (1) feedback on wireframes, and (2) perceptions regarding system implications. Our fourth level labels were organized in single point themes shown in Table \ref{tab:results}. These themes will be used to create a framework for future design iterations. 

\textbf{Results. }Overall, providers preferred interfaces that are simple yet engaging. As well, providing scientific context such as established literature in medicine on patient-centered communication may increase their trust and understanding of the metrics displayed. Familiarity with the interface was controversial, as some providers cautioned it may increase digital fatigue, while others believed it would simplify user interactions. Privacy concerns varied as some providers wanted their metrics to remain confidential due to fear of being reprimanded unfairly while others believed sharing them with colleagues and leadership may improve their performance. Providers also shared concerns on personal and institutional barriers that may interfere with the usage and adoption of our tools, such as lack of funding and time allocated by their employers for engaging with our ecosystem. 

% \begin{figure}
%     \centering
%     \includegraphics[scale=0.5]{images/ad.png}
%     \caption{The themes from our affinity diagram will be used to create a framework for future design iterations. }
%     \label{fig:ad}
% \end{figure}

\begin{table}[]
\begin{tabular}{|l|l|l|l|}
\hline
\multicolumn{1}{|c|}{\textbf{Theme}} & \multicolumn{1}{c|}{\textbf{Concerns}} & \multicolumn{1}{c|}{\textbf{Example}} & \multicolumn{1}{c|}{\textbf{Solution}} \\ \hline
Efficiency                           &   \makecell[cl]{Complex interface\\ Frequency of usage}            & \makecell[cl]{Displaying monthly metrics\\ on dashboard}                      & Simplify interface                                                \\ \hline
\makecell[cl]{Context-\\awareness}                     & \makecell[cl]{Personal\\ Scientific}                             & \makecell[cl]{Taking patient health \\\mbox{complexity} into \mbox{consideration}}          & \makecell[cl]{Displaying relevant citations \\ \mbox{for metrics}}       \\ \hline
Privacy                              & \makecell[cl]{Private\\ Shared within institution} & \makecell[cl]{Protecting personal data while\\ holding providers accountable} & \makecell[cl]{Sharing anonymized data \\within institution}                        \\ \hline
Barriers                             & \makecell[cl]{Institutional\\ Personal}                          & \makecell[cl]{Negative metrics may create\\ \mbox{defensiveness} among users}        & \makecell[cl]{Incorporating both negative and\\ positive metrics in our ecosystem} \\ \hline
\end{tabular}
\caption{A brief overview of our qualitative analysis. Each theme has two affiliated concerns, an example of the concerns highlighted in interviews, and a possible solution for future design iterations.}
\label{tab:results}
\end{table}

\section{Discussion and Next Steps}
This paper presents preliminary results from a first round of design iterations and feedback from primary care providers. Our work is ongoing and will continue to expand our understanding of how to deliver feedback to clinical providers on implicit bias in their communication with patients. We identified four pertinent themes through our analysis: efficiency, context-awareness, privacy, and addressing barriers. For future design iterations, we are recruiting primary care providers across the United States to gain a better representation of the population. Sessions with providers will follow a format similar to the one mentioned in this paper, with each design iteration guided by the feedback we receive from providers in the interviews. Based on the results of this initial study, we will specifically focus on simplifying our designs, making them less intrusive, and more compelling. After the conclusion of our design iterations, we will begin developing our ecosystem to communicate potential biases during patient-provider interactions.

\section{Acknowledgements}
We thank our colleagues Emily Bascom, Steven Rick, Lisa Dirks, Wanda Pratt, and Janice Sabin who provided helpful insight that greatly assisted with this work.

\bibliographystyle{ACM-Reference-Format}
\bibliography{sample-base}

\end{document}